\documentclass[slac_one]{revtex4}
\usepackage{graphicx}
\usepackage{fancyhdr}
\usepackage{amsmath,amssymb}
\newcommand{\sms}{\widetilde m}
\newcommand{\mGUT}{M_{\rm GUT}}

\newcommand{\GeV}{{\ensuremath\rm \, GeV}}
\newcommand{\TeV}{{\ensuremath\rm \, TeV}}

\newcommand{\fb}{{\ensuremath\rm \, fb}}
\newcommand{\ab}{{\ensuremath\rm \, ab}}
\newcommand{\pb}{{\ensuremath\rm \, pb}}

\begin{document}

\preprint{DESY 05--116}
\title{{\small{2005 International Linear Collider Workshop - Stanford,
U.S.A.}}\\ 
\vspace{12pt}
Split Supersymmetry at Colliders} 

%

\author{W.~Kilian}
\affiliation{Deutsches Elektronen-Synchrotron DESY, D--22603 Hamburg, Germany}

\author{T.~Plehn$^1$}
\footnotetext[1]{Heisenberg Fellow}
\affiliation{MPI f\"ur Physik, F\"ohringer Ring 6, D--80805 M\"unchen,
  Germany}

\author{P.~Richardson}
\affiliation{Institute for Particle Physics Phenomenology, University of
  Durham, DH1 3LE, UK}

\author{E.~Schmidt}
\affiliation{Fachbereich Physik, University of Rostock, D-18051
  Rostock, Germany}

\begin{abstract}
We consider the collider phenomenology of split-supersymmetry
models. Despite the challenging nature of the signals in these models
the long-lived gluino can be discovered with masses above $2\,\TeV$ at
the LHC. At a future linear collider we will be able to observe the
renormalization group effects from split supersymmetry on the
chargino/neutralino mixing parameters, using measurements of the
neutralino and chargino masses and cross sections.  This
indirect determination of chargino/neutralino anomalous Yukawa
couplings is an important check for supersymmetric models in general.
\end{abstract}

\maketitle

\thispagestyle{fancy}


\section{INTRODUCTION} 
Split supersymmetry~\cite{SpS} is a possibility to evade many of the
phenomenological constraints that plague generic supersymmetric
extensions of the Standard Model.  By splitting the
supersymmetry-breaking scale between the scalar and the gaugino
sector, the squarks and sleptons are rendered heavy (somewhere between
several $\TeV$ and the GUT scale), while charginos and neutralinos may
still be at the $\TeV$ scale or below.  This setup eliminates
dangerous flavor-changing neutral current transitions, electric dipole
moments, and spurious proton-decay operators without the need for mass
degeneracy between the sfermion generations.  The benefits of the
supersymmetry paradigm, in particular the unification of gauge groups
at a high scale and the successful dark-matter prediction, are
retained.

In the Higgs sector, the split-supersymmetry scenario requires a
fine-tuning that pulls the Higgs vacuum expectation value down to the
observed electroweak scale.  The extra Higgses of a supersymmetric
model are located at the sfermion mass scale.  This fine-tuning is
obviously unnatural, but it may (or may not) find a convincing
explanation in ideas beyond the realm of particle
physics~\cite{landscape,drees}.

In this talk (for more details, see~\cite{pheno}), we investigate the
split-supersymmetry scenario from a purely phenomenological point of
view.  We ask ourselves the question whether and how (i) the particles
present in the low-energy spectrum can be detected, and (ii) the
underlying supersymmetric nature of the model can be verified.  The
two tasks require combining LHC and ILC data.

The low-energy effective theory is particularly simple.  In addition
to the Standard Model spectrum including the Higgs boson, the only
extra particles are the four neutralinos, two charginos, and a gluino.
Since all squarks are very heavy, the gluino is long-lived.  The
gluino can be produced at the LHC only, while the charginos and
neutralinos can be accessible both at the LHC and at the ILC.

Renormalization group running without sfermions and heavy Higgses
lifts the light Higgs mass considerably above the LEP limit, solving
another problem of the MSSM.  Still, the Higgs boson is expected to be
lighter than about $200\GeV$.  Apart from this Higgs mass bound, the
only trace of supersymmetry would be the mutual interactions of
Higgses, gauginos, and Higgsinos, i.e., the chargino and neutralino
Yukawa couplings.  These couplings are determined by the gauge
couplings at the matching scale $\sms$, where the scalars are
integrated out.

\section{RENORMALIZATION GROUP EVOLUTION}
\begin{figure}[hbt]
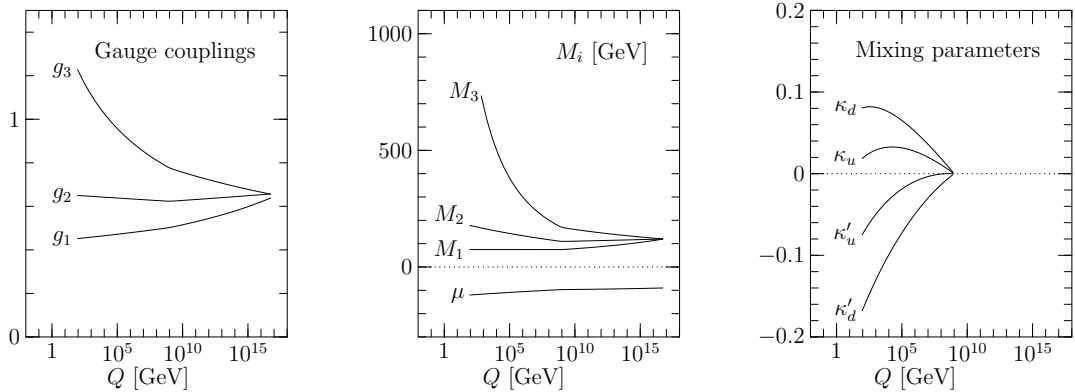

\begin{center}
\includegraphics[width=3.5cm]{rge-plots.1} \hspace{15mm}
\includegraphics[width=3.5cm]{rge-plots.2} \hspace{15mm}
\includegraphics[width=3.5cm]{rge-plots.3}
\end{center}
\vspace{3mm}
\caption{\label{fig:RGE} Renormalization group flow of the gauge
couplings (left), the gaugino--Higgsino mass parameters (centre), and
the anomalous gaugino--Higgsino mixing parameters defined in
eq.(\ref{eq:sps1_kappa}) (right). All curves are based on our reference point 
eq.(\ref{eq:sps1}).}
\end{figure}

Although, in the absence of sfermions, the overall phenomenology of
split-supersymmetry models does not depend very much on the particular
spectrum, for a quantitative analysis we have to select a specific
scenario.  To this end, we note that the popular assumption of
radiative symmetry breaking (i.e., a common scalar mass parameter for
sfermions and Higgses) has to be dropped, and the Higgsino mass
parameter $\mu$ is an independent quantity.  Assuming gauge
coupling unification and gaugino mass unification, we start from the
following model parameters at the grand unification scale
$\mGUT=6\times 10^{16}\GeV$:
\begin{align}
  M_1(\mGUT) = M_2(\mGUT) &= M_3(\mGUT) = 120\GeV,
&
  \mu(\mGUT) &= -90\GeV,
&
  \tan\beta  &= 4.
\label{eq:init}
\end{align}
For the SUSY-breaking scale we choose $\sms = 10^9\GeV$.
Figure~\ref{fig:RGE} displays the solutions of the renormalization
group equations with the input parameters set in
eq.(\ref{eq:init})~\cite{pheno}.  At the low scale $Q=m_Z$, we extract
the mass parameters:
\begin{align}
  M_1(Q=m_Z) &= 74.8\GeV & \qquad \qquad 
  M_3^{\overline{\text{DR}}}(Q=1 \TeV) &= 690.1\GeV \notag \\
  M_2(Q=m_Z) &= 178.1\GeV  & \qquad \qquad
  \mu(Q=m_Z) &= -120.1\GeV
\label{eq:sps1}
\end{align}
The resulting physical gaugino and Higgsino masses are:
\begin{align}
  m_{\tilde\chi^0_1} &= 71.1\GeV,
&   
  m_{\tilde\chi^0_2} &= 109.9\GeV, 
&
  m_{\tilde\chi^0_3} &= 141.7\GeV,
&
  m_{\tilde\chi^0_4} &= 213.7\GeV,
\notag\\
  m_{\tilde\chi^+_1} &= 114.7\GeV,
&
  m_{\tilde\chi^+_2} &= 215.7\GeV,
& 
  m_{\tilde g} &= 807\GeV
\label{eq:sps1_masses}
\end{align}
These mass values satisfy the LEP constraints.  Virtual effects of
a split supersymmetry spectrum on Standard Model observables have
recently been discussed in~\cite{Martin:2004id}.

The neutralinos
$\tilde\chi^0_{1,2,3,4}$ are predominantly bino, Higgsino, Higgsino,
and wino, respectively.  The Higgsino content of the lightest
neutralino is $h_f=0.2$, so the dark-matter condition~\cite{dm} is
satisfied.  To our given order the Higgs mass is $m_H=150\GeV$.

Because we integrate out the heavy scalars, the neutralino and
chargino Yukawa couplings deviate from their usual MSSM prediction,
parameterized by four anomalous Yukawa couplings~$\kappa$. We can
extract their weak-scale values from Fig.~\ref{fig:RGE}:
\begin{align}
  \frac{\tilde g_u}{g\sin\beta}   &\equiv 1 + \kappa_u
                                   = 1 + 0.018
 &\frac{\tilde g_d}{g\cos\beta}   &\equiv 1 + \kappa_d
                                   = 1 + 0.081 \notag \\ 
  \frac{\tilde g'_u}{g'\sin\beta} &\equiv 1 + \kappa'_u
                                   = 1 - 0.075
 &\frac{\tilde g'_d}{g'\cos\beta} &\equiv 1 + \kappa'_d
                                   = 1 - 0.17
\label{eq:sps1_kappa}
\end{align}

\section{LONG-LIVED GLUINOS}
Since the standard cascade decays of initial squarks and gluinos are
absent in this model, there are only two sources of new particles
left.  The gluino is produced in pairs from gluon-gluon and
quark-antiquark annihiliation.  Pairs of charginos and neutralinos are
produced through a Drell--Yan $s$-channel $Z$ boson, photon, or $W$
boson.  These cross sections are known to next-to-leading order
precision~\cite{LHC-NLO}.

\begin{figure}
\begin{center}
\includegraphics[width=5.5cm]{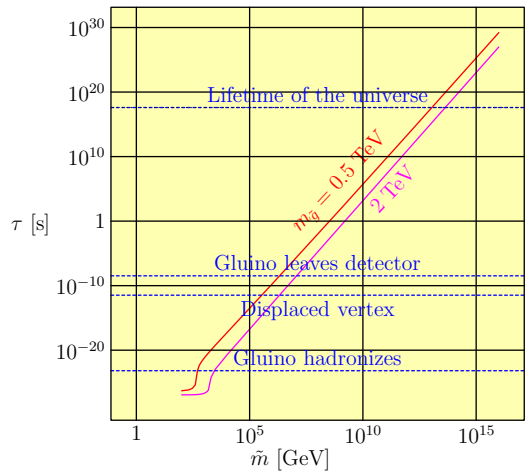}
\end{center}
\vspace{3mm}
\caption{\label{fig:gluino} Gluino lifetime~\cite{Muhlleitner:2003vg}
  as a function of the common scalar mass $\sms$.}
\end{figure}

Unless we have a-priori knowledge about the sfermion scale $\sms$, the
gluino lifetime is undetermined. Figure~\ref{fig:gluino} compares this
scale with other relevant scales of particle physics.  Once
$\sms\gtrsim 10^{3}\GeV$, the gluino hadronizes before decaying.  The
resulting states that consist of either of a gluino and pairs or
triplets of quarks, or of a gluino bound to a gluon, are called
$R$-hadrons~\cite{Rhadrons}.  For gluinos produced near
threshold, the formation of gluino-pair bound states (gluinonium) is
also possible and leads to characteristic signals~\cite{gluinonium}.

For $\sms>10^{6}\GeV$, the gluino travels a macroscopic distance.
If $\sms>10^{7}\GeV$, strange $R$-hadrons can also decay weakly, and
gluinos typically leave the detector undecayed or are stopped in the
material.  For even higher scales, $\sms>10^{9}\GeV$, $R$-hadrons
could become cosmologically relevant, since they affect
nucleosynthesis if their abundance in the early Universe is
sufficiently high~\cite{SpS}. 

If gluino decays can be observed, their analysis yields information
about physics at the scale $\sms$ and thus allows us to draw
conclusions about the mechanism of supersymmetry
breaking~\cite{gldecay}.

Without relying on gluino decays, there are two strategies for
detecting the corresponding $R$-hadrons~\cite{pheno,Hewett:2004nw}.
(i) The production of a stable, charged, $R$-hadron will give a signal
much like the production of a stable charged weakly-interacting
particle.  This signal consists of an object that looks like a muon
but arrives at the muon chambers significantly later than a muon owing
to its large mass.  (ii) While for stable neutral $R$-hadrons there
will be some energy loss in the detector, there will be a missing
transverse energy signal due to the escape of the $R$-hadrons. As
leptons are unlikely to be produced in this process, the signal will
be the classic SUSY jets with missing transverse energy signature.

\begin{figure}[hbt]
\begin{center}
\includegraphics[angle=90,width=0.35\textwidth]{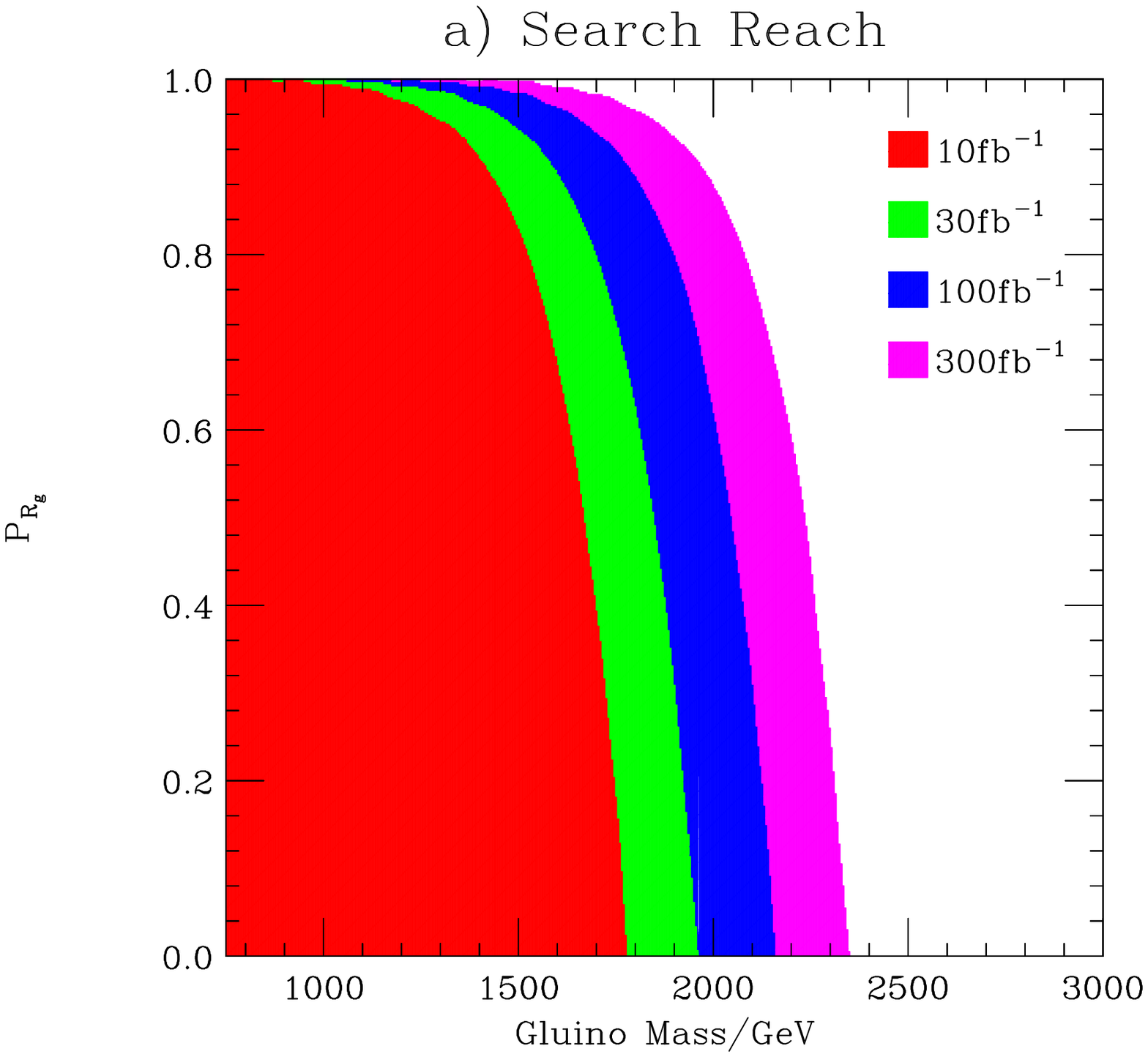}
\includegraphics[angle=90,width=0.35\textwidth]{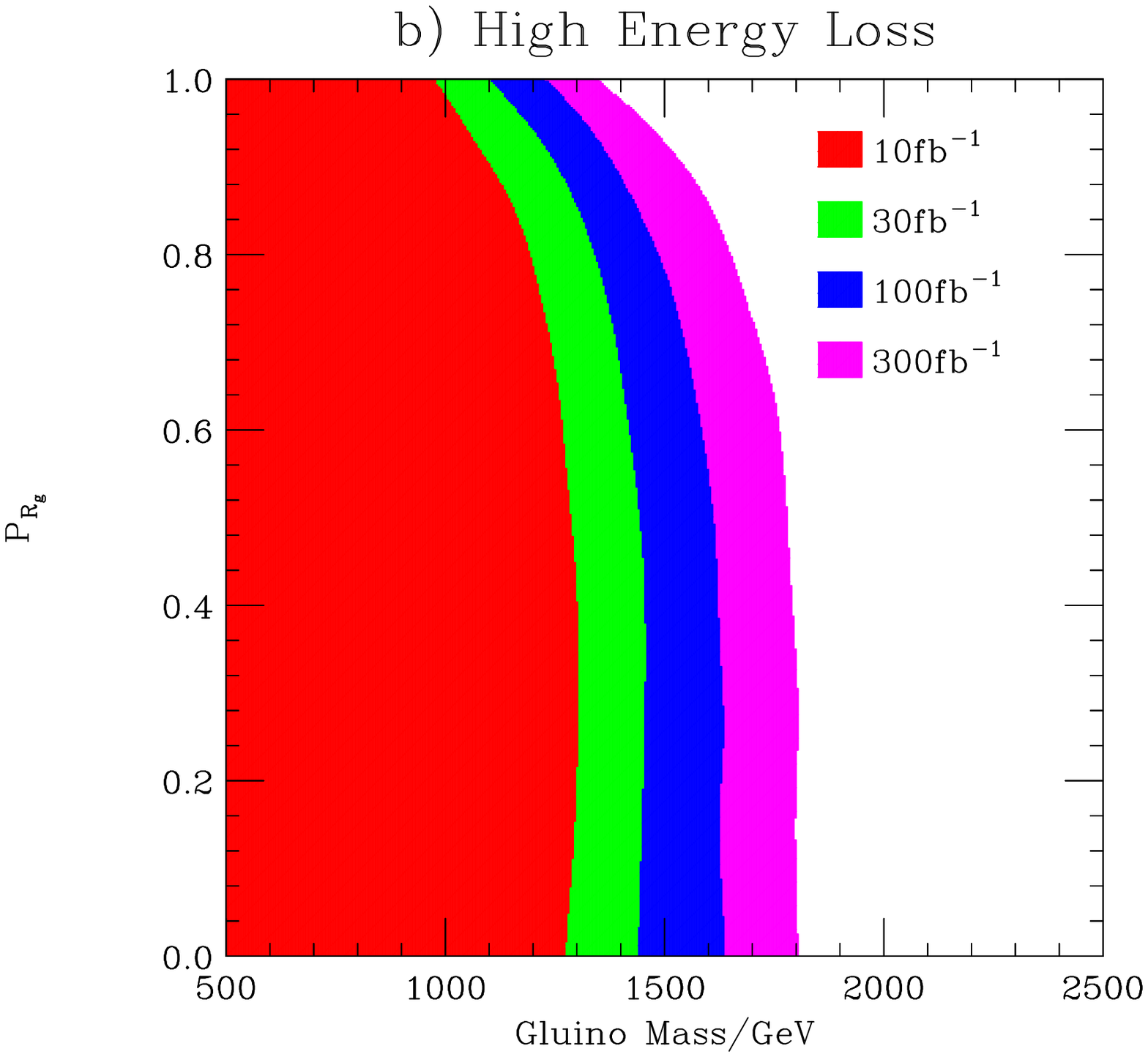}\\
\end{center}
\vspace*{-\baselineskip}
\caption{\label{fig:reach}
         Gluino discovery reach at the LHC for (a) charged
         $R$-hadrons, (b) neutral $R$-hadrons, from~\cite{pheno}.}
\end{figure}

In Fig.~\ref{fig:reach}, we show the expected discovery reach for both
channels, based on models for the $R$-hadron spectrum and the
$R$-hadron interaction in the detector that we have implemented in
HERWIG~\cite{HERWIG}.

\section{CHARGINO AND NEUTRALINO YUKAWA COUPLINGS}
If split supersymmetry should be realized in nature, the observation
of the gluino, charginos and neutralinos will only be the first
task. Once these states are discovered, we will have to show that at
the scale~$\sms$ they constitute a supersymmetric Lagrangian.

A quantitative trace of this is given by the off-diagonal elements in
the mass matrices that derive from the gaugino-higgsino-Higgs
couplings~(\ref{eq:sps1_kappa}). They determine the mixing of gauginos
and higgsinos into charginos and neutralinos as mass eigenstates.
Simultaneously, they also constitute the neutralino and chargino
Yukawa couplings.  In split supersymmetry, the renormalization flow
below the sfermion scale~$\sms$ induces non-zero values of order
$\kappa^{(\prime)}_i = -0.2\ldots 0.2$.  If we are able to detect
deviations of this size at a collider, we can both establish the
supersymmetric nature of the model and verify the matching condition
to the MSSM at $\sms$.

To measure the neutralino and chargino mixing matrices, a precise mass
measurement is necessary.  This is possible (for mass differences, at
least) at the LHC and, to a better accuracy, at the ILC. Without
gaugino--Higgsino mixing the mass matrices would be determined by the
MSSM parameters $M_1, M_2$ and $\mu$. The gaugino--Higgsino mixing
adds terms of the order of $M_Z$ and introduces the additional
parameter $\tan\beta$, leading to four MSSM parameters altogether.


In $e^+e^-$ collisions, for the parameter set~(\ref{eq:sps1}) almost
all chargino and neutralino production channels have cross sections
larger than $0.1\fb$, and the threshold value for
$\tilde\chi^+_1\tilde\chi^-_1$ production is as large as
$1\;\pb$~\cite{pheno}. The NLO electroweak corrections to these
production cross sections have been calculated in~\cite{ILC-NLO}. A
linear collider with moderate energy and high luminosity would be
optimal to probe all these processes, and some kind of fit is the
proper method to extract the weak-scale Lagrangian parameters.

\begin{figure}[t]
\begin{center}
\includegraphics[width=12.0cm]{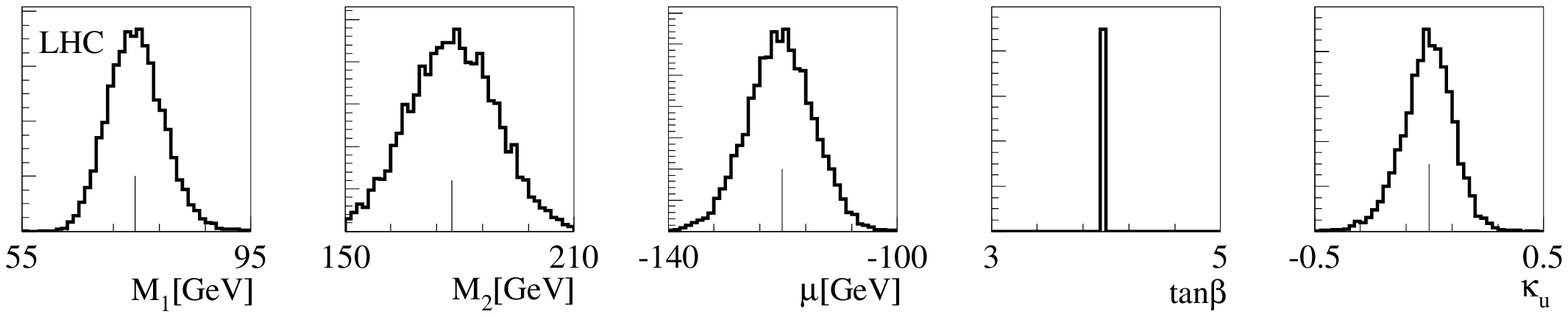} \\
\includegraphics[width=12.0cm]{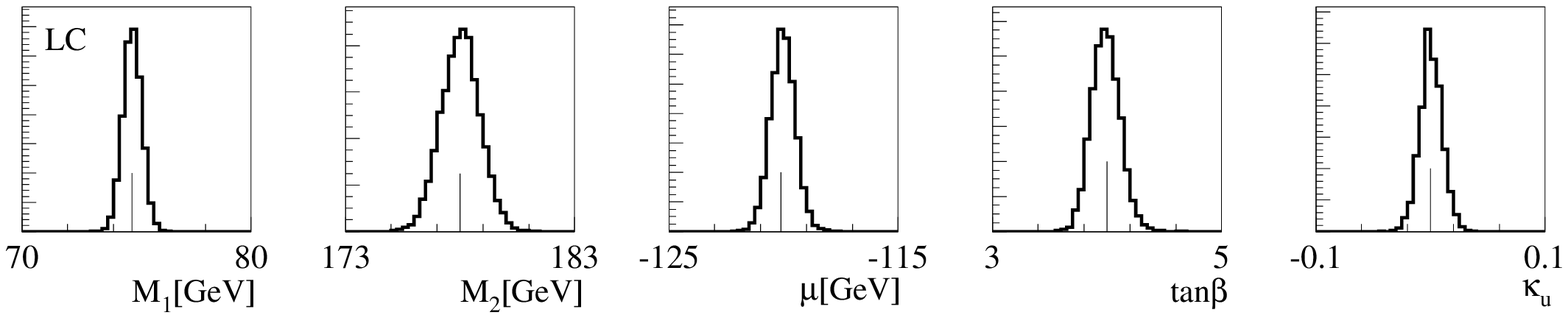} \\
\end{center}
\vspace{-8mm}
\caption{\label{fig:fit_1} Fit to 10000 sets of mass and cross section
pseudo-measurements at the LHC (upper) and at the ILC
(lower). The fitted parameters include only $\kappa_u$ with a central
value zero. At the LHC $\tan\beta=4$ is fixed.}
\end{figure}

\begin{figure}[t]
\begin{center}
\includegraphics[width=12.0cm]{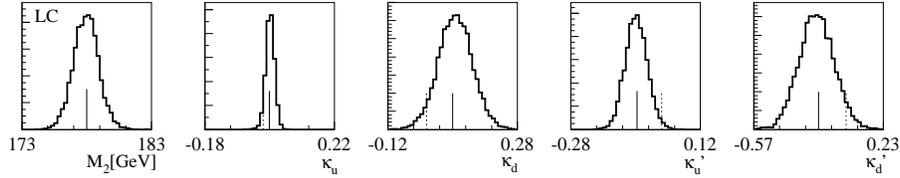}
\end{center}
\vspace{-8mm}
\caption{\label{fig:fit_2} Fit to 10000 sets of mass and cross section
pseudo-measurements at a future linear collider. All four
$\kappa^{(\prime)}_i$ are extracted simultaneously. The central values
are set to the example split supersymmetry values. The MSSM zero
prediction is indicated in the lower line of histograms.}
\end{figure}

We compute the masses and the cross sections for all pair-production
processes, with the exception of the $\tilde\chi^0_1 \tilde\chi^0_1$
channel. To all observables we assign an experimental error, which in
our simplified treatment is a relative error of $0.5\%$ on all
linear-collider mass measurements~\cite{Aguilar-Saavedra:2001rg},
$5\%$ on all LHC mass measurements~\cite{Bachacou:1999zb}, and the
statistical uncertainty on the number of events at a linear collider
corresponding to $100 \fb^{-1}$ of data at a $1 \TeV$ collider after
all efficiencies.

\begin{table}
\small
\begin{center}
\begin{tabular}{|l||c|c|c||r|r|r|r|}
\hline
          & Fit $\tan\beta$   & $m_i$    & $\sigma_{ij}$ & $\Delta \kappa_u$   & $\Delta \kappa_d$ & $\Delta \kappa'_u$  & $\Delta \kappa'_d$\\

\hline \hline
Tesla     &                   & $\bullet$ & $\bullet$      & $0.9 \times 10^{-2}$ & $3 \times 10^{-2}$ & $1.3 \times 10^{-2}$ & $4  \times 10^{-2}$ \\
Tesla     & $\bullet$          & $\bullet$ & $\bullet$      & $1.2 \times 10^{-2}$ & $5 \times 10^{-2}$ & $2   \times 10^{-2}$ & $5  \times 10^{-2}$ \\
Tesla     &                   & $\bullet$ &               & $1.1 \times 10^{-2}$ & $5 \times 10^{-2}$ & $3   \times 10^{-2}$ & $8  \times 10^{-2}$ \\
Tesla     & $\bullet$          & $\bullet$ &               & $1.2 \times 10^{-2}$ & $11\times 10^{-2}$ & $4   \times 10^{-2}$ & $8  \times 10^{-2}$ \\
LHC       &                   & $\bullet$ &               & $2.2 \times 10^{-1}$ & $6 \times 10^{-1}$ & $2.7 \times 10^{-1}$ & $8  \times 10^{-1}$ \\
\hline \hline
Tesla     &                   & $\bullet$ & $\bullet$      & $1.4 \times 10^{-2}$ & $5 \times 10^{-2}$ & $3   \times 10^{-2}$ & $10 \times 10^{-2}$ \\
Tesla & $\bullet$          & $\bullet$ & $\bullet$      & $1.7 \times 10^{-2}$ & $9 \times 10^{-2}$ & $4   \times 10^{-2}$ & $13 \times 10^{-2}$ \\
Tesla     & fix $\tan\beta=3$ & $\bullet$ & $\bullet$      & $1.6 \times 10^{-2}$ & $4 \times 10^{-2}$ & $4   \times 10^{-2}$ & $9  \times 10^{-2}$ \\
Tesla$^*$ & $\kappa_i \ne 0$  & $\bullet$ & $\bullet$      & $1.4 \times 10^{-2}$ & $5 \times 10^{-2}$ & $4   \times 10^{-2}$ & $11 \times 10^{-2}$ \\
\hline
\end{tabular}
\end{center}
\caption{\label{tab:fit} Error on the determination of $\kappa_i$
from measured masses and possibly production cross sections. For the
first five lines, all but one $\kappa$ are fixed to zero, the fitted
$\kappa$ has the central value zero. In the last four lines, all four
$\kappa_i$ are fitted simultaneously. The very last line assumes the
predicted central values of $\kappa_i$ in our parameter point.}
\end{table}

Around the central parameter point we randomly generate 10000 sets of
pseudo-measurements, using a Gaussian smearing. Out of each of these
sets we extract the MSSM parameters by a global fit method.  The fit
results (see Fig.~\ref{fig:fit_1}, top) show that at the LHC we can
extract the Lagrangian mass parameters with reasonable precision.
There is sensitivity to one Higgs-sector parameter, which we can take
either as $\tan\beta$ or as one of the mixing parameters.  If we fix
$\tan\beta=4$, the precision on $\kappa_u$ is sufficient to verify
consistency with a supersymmetric underlying theory.  However, using
LHC data alone, a simultaneous fit of all parameters gives only very
weak constraints on the anomalous Yukawa couplings
(Tab.~\ref{tab:fit}), so no conclusions about the split-supersymmetry
renormalization effects can be drawn.

The higher precision of measurements at the ILC, in particular adding
cross sections as independent observables, allows us to improve the
precision on a five-parameter fit (Fig.~\ref{fig:fit_1}, bottom) or to
simultaneously fit all Lagrangian parameters (Fig.~\ref{fig:fit_2}).
This is the proper treatment, unless we would have reasons to believe
that some of the $\kappa^{(\prime)}_i$ are predicted to be too small
to be measured.  The results for the precision in determining the
anomalous couplings are listed in Tab.~\ref{tab:fit}.  (Note that in a
complete fit, $\tan\beta$ is no longer an independent parameter, so we
can fix it to some given value.)

These results for the linear collider indeed indicate that we could
not only confirm that the Yukawa couplings and the neutralino and
chargino mixing follow the predicted MSSM pattern; for the somewhat
larger $\kappa'_i$ values we can even distinguish the complete
weak-scale MSSM from a split supersymmetry spectrum.


While the elements of the neutralino and chargino mixing matrices
depend on the Yukawa couplings in a complicated way, the cross
sections for chargino/neutralino pair production in association with a
Higgs boson are directly proportional to these parameters.  Decays of
the kind $\chi^\pm_2\to\chi^\pm_1H$ or $\chi^0_j\to\chi^0_iH$ would
carry the same information, but typically are kinematically forbidden
in split supersymmetry scenarios.

Associated production of charginos and neutralinos with a Higgs boson
in the continuum can in principle be observed at a high-luminosity
$e^+e^-$ collider.  The cross sections for some of these channels
exceed $0.1\;\fb$ with little background~\cite{pheno}, so with
$1\;\ab^{-1}$ of luminosity we would expect some events of this kind
to be detectable.  However, due to the small rates for these
processes, the achievable precision for parameter determination is
limited.  The measurement of masses and pair-production cross
sections will be the key for establishing split supersymmetry as the
underlying physical scenario.

\begin{acknowledgments}
W.K. is supported by the German Helmholtz-Gemeinschaft, Contract
No.\ VH--NG--005.
\end{acknowledgments}


\end{document}